\documentclass[iop,twocolappendix]{emulateapj}
\usepackage{apjfonts}
\usepackage{epstopdf}
\usepackage[caption = false]{subfig}
\usepackage{amsmath}
\usepackage{natbib}
\usepackage[backref,breaklinks,colorlinks,citecolor=blue]{hyperref}

\newcommand {\aplt} {\ {\raise-.5ex\hbox{$\buildrel<\over\sim$}}\ }

\newcommand{\MgII}{Mg{\sevenrm II}}
\newcommand{\OIII}{O{\sevenrm III}}
\usepackage{comment}
\font\sevenrm=cmr7 scaled 1000

\begin{document}

\title{The GALEX Time Domain Survey. II.  Wavelength-Dependent Variability of Active Galactic Nuclei in the Pan-STARRS1 Medium Deep Survey}
\author{T. Hung\altaffilmark{1}, S. Gezari\altaffilmark{1}, D. O. Jones\altaffilmark{2}, R. P. Kirshner\altaffilmark{3}, R. Chornock\altaffilmark{4}, E. Berger\altaffilmark{3}, A. Rest\altaffilmark{5}, M. Huber\altaffilmark{6}, G. Narayan\altaffilmark{7}, D. Scolnic\altaffilmark{8}, C. Waters\altaffilmark{6}, R. Wainscoat\altaffilmark{6}, D. C. Martin\altaffilmark{9}, K. Forster\altaffilmark{9}, and J. D. Neill\altaffilmark{9}}
\altaffiltext{1}{Department of Astronomy, University of Maryland, College Park, MD 20742, USA}
\altaffiltext{2}{Department of Physics and Astronomy, Johns Hopkins University, Baltimore, MD 21218, USA}
\altaffiltext{3}{Harvard-Smithsonian Center for Astrophysics, 60 Garden St., Cambridge, MA 02138, USA}
\altaffiltext{4}{Astrophysical Institute, Department of Physics and Astronomy,
251B Clippinger Lab, Ohio University Athens, OH 45701,
USA}
\altaffiltext{5}{Space Telescope Science Institute, 3700 San Martin Drive,
Baltimore, MD 21218, USA}
\altaffiltext{6}{Institute for Astronomy, University of Hawaii, 2680 Woodlawn Drive, Honolulu, HI 96822, USA}
\altaffiltext{7}{National Optical Astronomy Observatory, 950 North Cherry Avenue, Tucson, AZ 85719, USA}
\altaffiltext{8}{Kavli Institute for Cosmological Physics, University of Chicago, Chicago, IL 60637, USA}

\altaffiltext{9}{Cahill Center for Astrophysics, California Institute of Technology, 1216 East California Boulevard, Mail Code 278-17, Pasadena, California 91125, USA}

\begin{abstract}
We analyze the wavelength-dependent variability of a sample of spectroscopically confirmed active galactic nuclei (AGN) selected from near-UV ($NUV$) variable sources in the \textsl{GALEX} Time Domain Survey that have a large amplitude of optical variability (difference-flux S/N $>$ 3) in the Pan-STARRS1 Medium Deep Survey (PS1 MDS). By matching \textsl{GALEX} and PS1 epochs in 5 bands ($NUV$, $g_{P1}$, $r_{P1}$,
$i_{P1}$, $z_{P1}$) in time, and taking their flux difference, we create co-temporal difference-flux spectral energy distributions ($\Delta f$SEDs) using two chosen epochs for each of the 23 objects in our sample on timescales of about a year. 

We confirm the "bluer-when-brighter" trend reported in previous studies, and measure a median spectral index of the $\Delta f$SEDs of $\alpha_{\lambda}$ = 2.1 that is consistent with an accretion disk spectrum.  We further fit the $\Delta f$SEDs of each source with a standard accretion disk model in which the accretion rate changes from one epoch to the other. In our sample, 17 out of 23 ($\sim$74\%) sources are well described by this variable accretion-rate disk model, with a median average characteristic disk temperature $\bar{T}^*$ of $1.2\times 10^5$~K that is consistent with the temperatures expected given the distribution of accretion rates and black hole masses inferred for the sample. Our analysis also shows that the variable accretion rate model is a better fit to the $\Delta f$SEDs than a simple power law.
\end{abstract}


\section{Introduction}
Active galactic nuclei (AGN) are known to vary across the observable electromagnetic spectrum on timescales ranging from seconds to years. Large time-domain surveys have confirmed variability as a ubiquitous
characteristic of AGN \citep[e.g.][]{Vandenberk04,Wilhite05}, ruling out models
involving extrinsic factors such as gravitational microlensing, star
collisions or multiple supernovae or starbursts near the nucleus \citep[][and references therein]{Kokubo15}.
In particular, the origin of UV/optical variability is of great interest for its connection to the AGN central engine, since the UV/optical continuum is thought to arise directly from the accretion disk around a supermassive black hole. 

A well-established characteristic of AGN UV/optical variability is the
"bluer-when-brighter" trend, in which the source is bluer in the bright state
than in the faint state \citep{Vandenberk04,Wilhite05,Schmidt12}. (Although a
reverse trend was observed for low-luminosity AGN in \citet{2016ApJ...821...86H}).  However, the interpretation of the "bluer-when-brighter" trend is still under debate. Whether or not the intrinsic AGN color becomes bluer
when it brightens, or if the red color in a low state is caused by
contamination from a non-variable component (e.g., host galaxy flux)
\citep{Winkler92}, cannot be distinguished due to the difficulties of
separating host galaxy light from the AGN.

\cite{2009ApJ...698..895K} found that the optical variability of an AGN is well described by a damped random walk (DRW) process, with model-fit timescales that are consistent with the thermal timescale of an accretion disk.  A similar conclusion was reached by \cite{2010ApJ...721.1014M} with a sample of $\sim$9000 spectroscopically confirmed quasars in SDSS Stripe 82. Furthermore, \cite{2010ApJ...721.1014M} found that the variability amplitude was anti-correlated with Eddington ratio, implying that the mechanism driving the optical variability was related to the accretion disk.

Several prevailing models for the origin of UV/optical variability in the AGN accretion disk
include thermal reprocessing of X-ray emission
\citep{1991ApJ...371..541K}, changes in the mass accretion rate
\citep[e.g.][]{Pereyra06,Li08}, and an extremely inhomogeneous accretion disk
\citep{Dexter11,Ruan14}. In an effort to examine these models, several studies have
been carried out.  For example, \cite{Ruan14} found that a
relative spectral variability composite spectrum constructed from SDSS quasars is better fit with an inhomogeneous disk model consisting of multiple
zones undergoing independent temperature fluctuations, than simply changes in a steady-state mass accretion rate. While the localized
temperature fluctuations may arise from magnetorotational instabilities (MRI)
in an accretion flow \citep[see discussions in][]{2009ApJ...704..781H,2012A&A...540A.114J,2013ApJ...778...65J}, it has also been pointed out by \cite{Kokubo15} that
the inhomogeneous disk model predicts a weak inter-band correlation, which is
contradictory to what is often observed in the SDSS quasar light curves.
Kokubo (2015) attributed the successful fitting result in \cite{Ruan14} to the
use of a composite difference spectrum, in which the superposition of
localized flares at different radii smears out the inter-band flux-flux
correlation in individual quasars.

On the other hand, \citet[][ hereafter P06]{Pereyra06} presented the first
successful result of fitting a composite difference spectrum of SDSS quasars
to a standard thin disk model \citep{Shakura73} with changes in accretion rate
from one epoch to the next. Recently, many works that try to explain AGN
variability have also reached conclusions that are supportive of this variable
accretion rate model \citep{Li08,Sakata11}. The model suggests that the "bluer
when brighter" trend is caused by intrinsic AGN spectral hardening. The same
conclusion was reached in \cite{Wilhite05}, where they found steeper spectral
index in the composite flux difference spectrum than that in the composite
spectrum. Despite the elegance of this model, there is a discrepancy between
the AGN UV-optical variability timescale and the sound crossing and the viscous
timescales (by a factor of 10$^3$, see discussion in \autoref{subsec:timescales}) of the accretion disk. Furthermore, \cite{Schmidt12,2014ApJ...783...46K} reported that on the timescales of years, characteristic optical color variability in
individual SDSS stripe 82 quasars is larger than the color variability
predicted by a steady-state accretion disk with a varying accretion rate.

In this paper we continue the investigation of the origin of UV/optical
variability in quasars with several improvements. We apply the variable
accretion rate model (P06) to the difference-flux spectral energy
distributions (SEDs) of individual quasars instead of using a composite quasar
difference spectrum, and over a wavelength range broader than previous optical
studies ($\approx 1700-9000 \AA$ in observer's frame) thanks to the nearly simultaneous near-UV ($NUV$) observations from
\textsl{GALEX}. The sample of AGNs and quasars investigated in this paper is
selected from NUV-variability from the \textsl{GALEX} Time Domain Survey (TDS)
and is analyzed using broad-band \textsl{GALEX} $NUV$ and Pan-STARRS1 Medium
Deep Survey (PS1 MDS) optical $griz$ photometric data. In
\autoref{sec:selection}, we describe the sample selection and the
spectroscopic data used in this paper. We describe details of the P06 model in
\autoref{sec:model}. The creation of $\Delta f$SEDs, the spectral
fitting procedures, and the derivation of fundamental AGN parameters are
detailed in \autoref{sec:analysis}. We present our result in
\autoref{sec:results}. The implications of the results are discussed in
\autoref{sec:discuss}.

\section{Sample Selection and Data Reduction}
\label{sec:selection}
Throughout the analysis in this paper, we define an AGN as an optical
stochastic variable source with extended host morphology, while quasars are
defined as optical point sources with the following colors \citep{Gezari13}.

\begin{eqnarray}
u-g < 0.7 \\
-0.1 < g-r < 1.0 \nonumber 
\end{eqnarray}

The optical morphology classification and colors are obtained from CFHT $u$
band and the PS1 deep stack catalog ($griz$) in \cite{2016ApJ...821...86H}. The
classification of stochastic variablility was carried out in \cite{Kumar15} by
comparing the goodness of fit of the light curves to the Ornstein-Uhlenbeck process, a general version of the DRW, and the light curve shapes characteristic of supernovae.

We cross-match the \textsl{GALEX} TDS UV-variable source
catalog with the PS1 transient alert database with a search
radius of 2$\arcsec$ to obtain a sample of 335 UV and optical variable sources. 
For the analysis, we select a subsample of 24 large-amplitude optically
variable sources with a difference-flux signal-to-noise (S/N) $> 3$ in the $g$ band, for all of which we have spectra.  All but one source in this subsample are spectroscopically classified as active galaxies, with 11 classified as AGN and 12 classified as quasars using our definitions described above. Spectroscopy shows that the exception is a normal galaxy with unusual UV variability (see \autoref{subsec:exception} for discussion).  
\subsection{\textsl{GALEX} TDS}

\textsl{GALEX} TDS \citep{Gezari13} covered  $\sim$40 deg$^2$ of the sky and detected over 1000
UV variable sources with observations that span a 3-year baseline  with a cadence of 2 days.
\textsl{GALEX} TDS fields were designed to monitor 6 out of 10 PS1 MDS fields,
with 7 \textsl{GALEX} TDS pointings at a time to cover the 8 deg$^2$ field of
view of PS1. The typical exposure time per epoch is 1.5 ks (or a 5$\sigma$
point-source limit of $m_{AB}$ $\sim$ 23.3~mag). The \textsl{GALEX} far-UV (FUV)
detector became non-operational during the time of the full survey, and so
only near-UV ($NUV$) data is available for analysis. 

Sources were classified as $NUV$ variable in the \textsl{GALEX} TDS catalog if
they have at least one epoch in which $|m_k - \bar{m}|>5\sigma(\bar{m}, k)$
\citep{Gezari13}, where $m_k$ is the magnitude of epoch $k$ and $\bar{m}$ is calculated only from epochs that have a magnitude above the detection limit of
that epoch. The variable sources were then classified using a
combination of optical host colors and morphology, UV light curves, and
matches to archival X-ray and spectroscopy catalogs. The final sample of 1078
NUV variable sources consists of 62\% active galaxies (AGN and quasars), 10\%
variable stars (RR Lyrae, M dwarfs, and cataclysmic variables), and the rest
are sources without classification (generally without spectra).

\subsection{PS1 MDS}

The Pan-STARRS1 Medium Deep Survey (PS1 MDS) surveyed 10 spatially separated
fields, each with a 8 deg$^2$ field of view, with a set of 5 broadband
filters: $g_{P1}$, $r_{P1}$, $i_{P1}$, $z_{P1}$, $y_{P1}$ \citep{Kaiser2010}. The average cadence
was 6 epochs in 10 days with a 5$\sigma$ depth of m$_{AB}$$\sim$23.1 mag, 23.3~mag, 23.2~mag, 22.8~mag in $g_{P1}$, $r_{P1}$, $i_{P1}$, and $z_{P1}$, respectively \citep{Rest14}.
Each night, 5 MD fields were observed with a 3-day staggered cadence in
each filter with the following pattern: $g_{P1}$ and $r_{P1}$ in the same
night (dark time), followed by $i_{P1}$ and $z_{P1}$ on the subsequent second
and third night, respectively. During the week of the full moon, the $y_{P1}$
band was observed exclusively. We do not use the $y_{P1}$ band in our analysis
due to the large time interval between $y_{P1}$ and the other bands.

We use data from the transient alert system of PS1 MDS that is designed to
detect transient events. The images taken in PS1 MDS are detrended by the
Image Processing Pipeline (IPP) \citep{Magnier06}. Typically 8 dithered images
are taken per filter and stacked in a given night. Image subtraction of a template created from observations of two earlier epochs
from the nightly stacked image is performed with the \texttt{photpipe} pipeline
\citep{Rest05}. The system determines the spatially varying convolution kernel
in a robust way before performing a subtraction of the two images. A transient
detection is flagged if it shows positive detections with a signal-to-noise ratio (S/N)~$\geq$4 in at least three $griz_{P1}$ images within a time window of 10 days.
Note that most previous studies of AGN variability have used samples of
point-like quasars but avoided AGN with resolved galaxy hosts because the
host galaxy light is hard to separate from the AGN light. With difference
imaging, the non-variable component can be removed effectively, leaving only
the variable component of the AGN.

\subsection{Optical spectra}

We gather spectra for all the objects in our sample, including 5 from the first data release of the Sloan Digital Sky Survey \citep[SDSS DR1;]{2003AJ....126.2081A}, 8 from the archival Baryon Oscillation Spectroscopic Survey \citep[BOSS;][]{2013AJ....145...10D}, 1 from the archival 2dF Galaxy Redshift Survey (2dFGRS) \citep[2dFGRS;][]{2003astro.ph..6581C}, and the remaining 10 from the PS1 transients follow-up programs (PI Kirshner and PI Berger) using MMT Hectospec. 

The SDSS spectra were observed with a 2.5m telescope and a pair of fiber-fed double spectrographs at the Apache Point Observatory. Each SDSS spectrum is measured from 3800 \AA to 9200 \AA~on 2048 $\times$ 2048 CCDs with a resolving power ($\lambda/\Delta \lambda$) of $\sim$ 2000.

Hectospec is a multi-fiber spectrograph mounted on MMT that is capable of observing 300 objects simultaneously \citep{2005PASP..117.1411F}. During the observation, a 270 (lpmm) grating was used, which corresponds to a dispersion of 1.21 \AA/pix near 5000 \AA. The spectral coverage is from 3650 to 9200 \AA.

The SDSS spectra were extracted and calibrated with its own \texttt{spectro2d} pipeline while the MMT spectra were reduced with the automated \texttt{HSRED} pipeline, which is an \texttt{IDL} package based on the reduction routines of SDSS spectroscopic data. The optical spectra were reduced following a conventional reduction protocal that includes bias removal, flat fielding, cosmic ray removal, sky subtraction, 1-d spectrum extraction, and wavelength and flux calibration.

The 2dFGRS is a large spectroscopic survey conducted using the Two-Degree Field (2dF) multi-fiber spectrograph at the Anglo-Australian telescope. The survey spectra cover the wavelength range of 3600-8000~\AA~with a resolution of 9~\AA. Due to the lack of flux calibration files for the 2dFGRS spectra, we can only approximate the relative flux by dividing the 2dFGRS spectra by a response curve created from observing a set of stars with SDSS $ugriz$ photometry \citep{2002ApJ...569..582B}.

\section{Accretion Disk Model}
\label{sec:model}

\cite{Pereyra06} successfully modeled a composite residual spectrum of over
300 SDSS quasars with a standard disk model in which the mass accretion rate
changes from one epoch to another. In this paper, we attempt to take this
further by fitting $\Delta f$SEDs with the P06 model on an object per
object basis.

In the standard geometrically thin and optically thick accretion
disk model \citep{Shakura73}, the potential energy released as the
mass accretes onto a black hole is emitted locally. Considering the interplay
between mass moving inward and shear stresses transporting angular momentum
outward, the disk temperature profile can be described by

\begin{equation}
T(r)= T^*\left(\left(\frac{r_i}{r}\right)^3\left[1-\left(\frac{r_i}{r}\right)^{1/2}\right]\right)^{1/4},
\end{equation}
where $r_i$ is the inner disk radius and T$^*$ is the characteristic
temperature defined by

\begin{equation}
T^* = \left(\frac{3GM_{\text{bh}}\dot{M}_{\text{accr}}}{8\pi r_i^3\sigma}\right)^{1/4},
\end{equation}
where $G$ is the gravitational constant, $M_{\text{bh}}$ is the black hole
mass, and $\dot{M}_{\text{accr}}$ is the mass accretion rate.

If we assume the inner disk radius $r_i$ to be the radius of the innermost
stable circular orbit ( $R_\text{ISCO}$ = $6GM_{bh}/c^2$) around a
Schwarzschild black hole, the characteristic temperature $T^*$ becomes

\begin{equation}\label{eq:Tch}
T^* = \left(\frac{\dot{M}_{\text{accr}}c^6}{576\pi G^2M_{\text{bh}}^2\sigma_s}\right)^{1/4}.
\end{equation} 
Since $M_{\text{bh}}$ should not change significantly on a timescale of years,
changes in $T^*$ trace changes in accretion rate.

For a given temperature profile, flux from the disk is given by the sum over
blackbody spectra of annuli. The total flux from both sides of the disk is
given by

\begin{equation}
f_\lambda(T^*) = \int\displaylimits_{r_i}^{\infty} { \pi \, { {2 h c^2 \over \lambda^5} \over \exp \left( h c \over \lambda \, k T^* t(r/r_i) \right) - 1 } \, 4 \pi r \, dr },
\end{equation}
where $c$ is the speed of light, $h$ is Planck's constant, $k$ is Boltzmann's
constant and the function $t(x) = x^{-3}(1 - x^{-1/2})^{1/4}$. Define

\begin{equation}
s\equiv \frac{r}{r_i}.
\end{equation}
We can rewrite the expression for flux by normalizing the disk radius to the inner disk radius to:
\begin{equation}\label{eq:integral}
f_\lambda(T^*) = r_i^2 \int\displaylimits_{1}^{\infty} { \pi \, { {2 h c^2 \over \lambda^5} \over \exp \left( h c \over \lambda \, k T^* t(s) \right) - 1 } \, 4 \pi s \, ds }.
\end{equation}
The observed disk flux can be expressed as
\begin{equation}
f_{o\lambda} =  c_o \; g_\lambda(T^*),
\end{equation}
where $c_o$ is a constant that depends on the black hole mass (see definition of $R_\text{ISCO}$), the object's cosmological distance, and the disk viewing angle and $g_\lambda(T^*)$ is defined as the integral in \autoref{eq:integral}.
\begin{equation} \label{eq:model}
g_\lambda(T^*) \equiv \int\displaylimits_{1}^{\infty} { \pi \, { {2 h c^2 \over \lambda^5} \over \exp \left( h c \over \lambda \, k T^* t(s) \right) - 1 } \, 4 \pi s \, ds }.
\end{equation}

The P06 model assumes the disk evolved from one steady state to the other between two arbitrary epochs with a change in mass accretion rate $\Delta \dot{M}_{\text{accr}}$. The observed flux difference is then
\begin{equation} \label{eq:solve}
\Delta f_{o\lambda} =  c_o (g_\lambda(T_2^*)-g_\lambda(T_1^*)),
\end{equation}
with a mean characteristic temperature that is defined as
\begin{equation}
\bar{T}^* = \frac{1}{2}(T_2^*+T_1^*),
\end{equation}
and

\begin{equation}
\Delta T^* = T_2^*-T_1^*.
\end{equation}
P06 performed Taylor expansion about $\bar{T}^*$ for g$_\lambda$(T$_1^*$) and g$_\lambda$(T$_2^*$) and reduced \autoref{eq:solve} to its primary term
\begin{equation}\label{eq:model_tay}
  \Delta f_{o\lambda}
\approx  c_o \, \Delta T^* \, \left. \partial g_\lambda \over \partial T^* \right|_{\bar{T}^*}.
\end{equation}
One thing to note here is that $T_1^*$ and $T_2^*$ cannot be determined
individually. Because the partial derivative term here is a function of
$\bar{T}^*$, one would expect to find pairs of $T_1^*$ and $T_2^*$ that have
the same $\bar{T}^*$ corresponding to several local minima in $\chi^2$ space.
To avoid problems caused by this degeneracy, we only fit the mean
characteristic temperature $\bar{T}^*$ in this work.

\section{Analysis}
Our analysis consists of three steps: (1) Selecting a sample of objects with a bright and a faint phase that have synchronous interband observations for each object, (2) Constructing the difference-flux
SED using the two chosen epochs, and (3) fitting the $\Delta f$SEDs with the P06 model
(\autoref{eq:model_tay}). The free parameters in the P06 model are
$\bar{T}^*$ and an arbitrary normalization constant.

\subsection{Bright and faint phase}
Cross matching \textsl{GALEX} $NUV$ and PS1 $griz$ data for studying time variable phenomena can be tricky since the observations were not made at the same time for different telescopes and filters. To preserve the simultaneity of the data, we first group each GALEX epoch with PS1 by minimizing the interband observation gap with respect to \textsl{GALEX} $NUV$. For each object, the two best aligned \textsl{GALEX} epochs with $griz$ observations within 9 days of the $NUV$ observation are chosen as their bright and faint epochs, where bright and faint are decided by the $NUV$ magnitude.

\begin{figure}
\centering
\includegraphics[width=3.5in, angle=0]{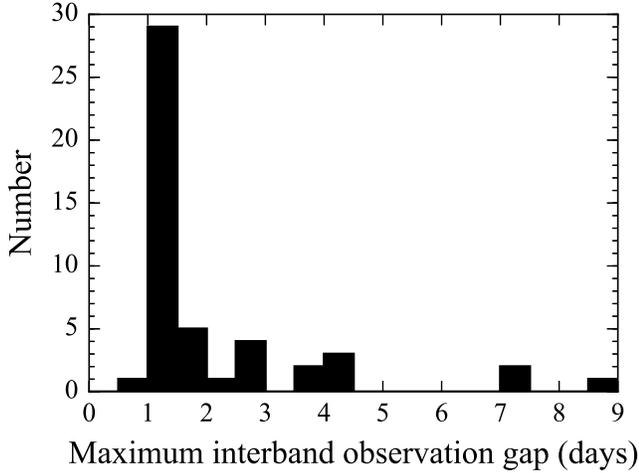}
\caption{Distribution of the maximum obervation gap between $NUV$ and the optical bands of the two epochs used to construct $\Delta f$ SEDs in our sample.}
\label{fig:gap}
\end{figure}

Next, we take the flux difference between the bright and the faint epochs for each object and propagate the uncertainties to calculate the S/N for the difference-flux ($\Delta f$). 
We exclude the small optical variability epochs with a $\Delta f$ S/N lower than 3 in $g$ band.
Among the 335 objects that have overlapped in both \textsl{GALEX} and PS1 surveys, we found a sample of 24 objects that have more than one co-temporaneous observations between all 5 bands and have large amplitude of variability (S/N($\Delta f_g$) $>$ 3) between these two chosen epochs. \autoref{fig:gap} shows the distribution of the largest observation gaps in each faint and bright epoch for all 24 objects. Most of the epochs used for our analyses have nearly simultaneous observations across $NUV$ and optical bands.

\label{sec:analysis}
\subsection{Construction of the difference-flux SEDs}
\label{subsec:sed}

%%% BEGIN METHOD

%We first find all epochs with nearly simultaneous $NUV$ and optical
%observations for each source in the PS1 transient alert database, where we require observations in at least three optical bands from within 8 days of the $NUV$ observation, and label the
%epoch with the faintest $NUV$ flux as the "reference". 

Once the bright and faint epochs are determined, we reprocess the image subtraction to obtain more accurate flux measurements at the two chosen epochs using a customized pipeline developed by Jones et al. (in prep). Unlike the templates used in \texttt{photpipe} that are created from stacking observations of two earlier epochs resulting in
double the typical exposure time of the science epoch, the templates in Jones et al. (in prep) stack images from every season except the season the event was discovered in ($\sim$ 70 images).  
This pipeline then performs forced photometry on stars in each field to determine the zeropoints of each image after determining accurate stellar centroids using average positions from over 300 individual images. The new pipeline also increased the number of stars used to build the PSF model to improve the fit and reduce the photometric noise.

It is worth noting that the
\textsl{GALEX} $NUV$ magnitudes are aperture magnitudes extracted using a 6" radius aperture that include the host
flux, while the PS1 $g_{\rm P1} r_{\rm P1} i_{\rm P1} z_{\rm P1}$ magnitudes are measured after difference imaging,
where flux from an earlier template has been subtracted off. We did not perform image differencing on GALEX images since most galaxies are unresolved by the \textsl{GALEX} NUV 5".3 FWHM point-spread function (PSF). When we
construct a $\Delta f$SED, we subtract the flux in each band in the
faint epoch from the bright epoch in flux space, so that we are only
sensitive to the SED of the change in flux ($\Delta f$) between the two epochs, and we are insensitive to which template was used
to construct the PS1 difference magnitude. 
%as long as the same template was used for both epochs. 

%We then calculate the difference-flux S/N for each GALEX epoch with nearly simultaneous optical observations from PS1 and exclude the small optical variability epochs with a difference-flux S/N lower than 3 in the $g$ band. One representative "bright" epoch was chosen from the large amplitude epochs for each source that
%has the most number of filters that are best aligned in time. 

\begin{figure*}
\centering
\includegraphics[width=6in, angle=0]{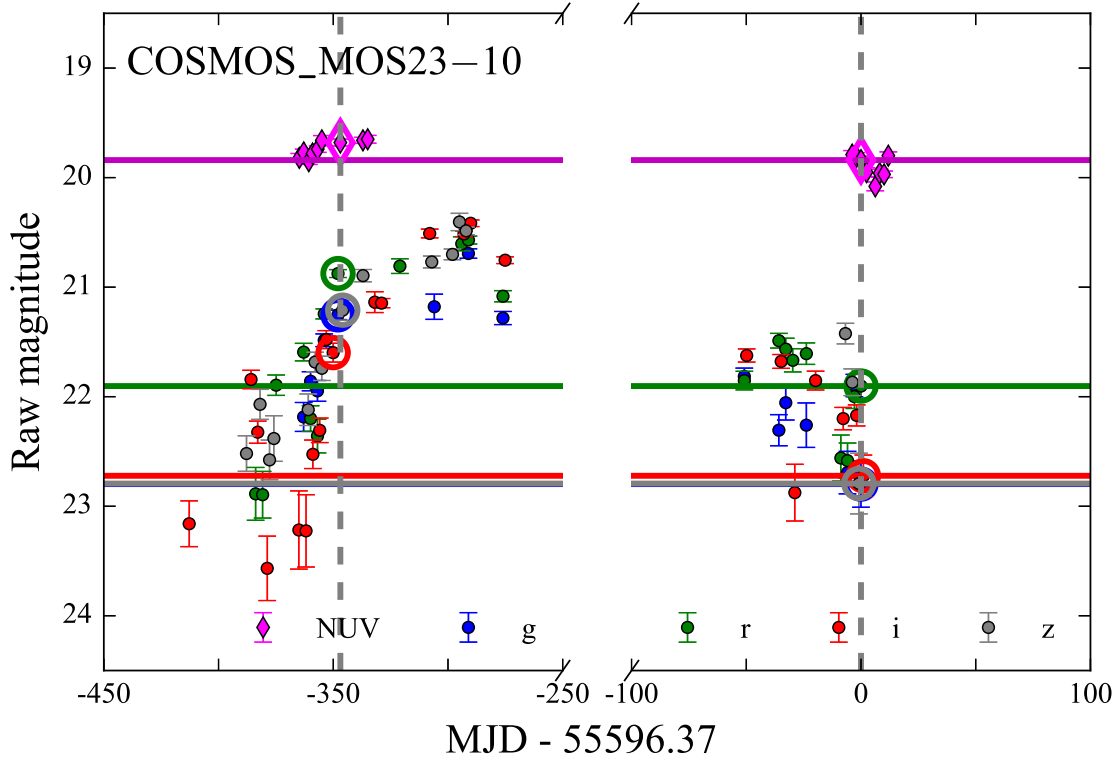}
\caption{The light curve of COSMOS\_MOS23-10 with epochs shown in time with respect to the
reference epoch. Dashed lines mark the \textsl{GALEX} epochs of a bright and faint
(reference) state. The y-axis shows the AB magnitude of the NUV/optical data
in its original form. For \textsl{GALEX} NUV data, the raw magnitude includes
flux from the host galaxy while the PS1 optical points show the magnitudes of
the template subtracted flux. Data circled is used in creating difference-flux
SED. The raw magnitude in PS1 is shown as the brightness of a difference image
while in
\textsl{GALEX} it is the brightness of the given epoch. The horizontal lines
mark the raw magnitude of the reference with different colors indicating each
band.}
\label{fig:lc}
\end{figure*}

The flux subtraction is performed in
flux density units (erg s$^{-1}$ cm$^{-2}$ \AA$^{-1}$) to match the model
prediction in f$_\lambda$~$-$~$\lambda$ space. We demonstrate the construction of $\Delta f$SEDs in \autoref{fig:lc}, which shows the
light curve of COSMOS\_MOS23-10, with vertical dashed lines marking the two \textsl{GALEX} epochs that correspond to the faint and bright phase.  The horizontal lines denote the reference magnitude of each filter at the faint epoch.

\begin{figure}
\centering
\includegraphics[width=3.5in, angle=0]{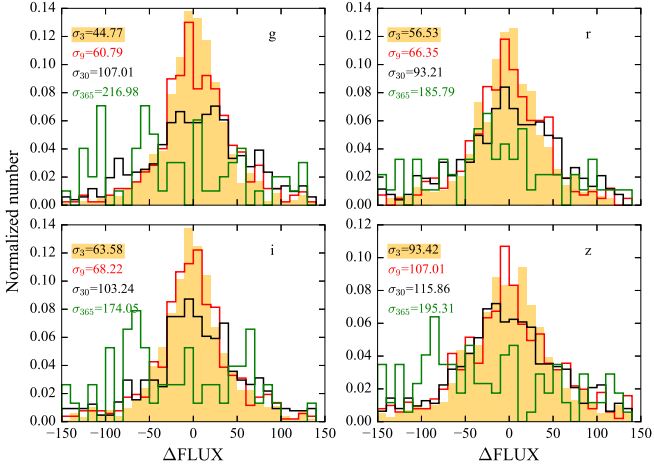}
\caption{The distribution of flux difference in 4 different $\Delta t$ bins: 3$\pm$0.5d (orange filled), 9$\pm$0.5d (red), 30$\pm$0.5d (black), 365$\pm$0.5d (green) in PS1 $griz$ bands. The flux used here corresponds to a zeropoint of 27.5. The histograms are normalized by the total number of pairs in each $\Delta t$ bin.}
\label{fig:timescale_sigma}
\end{figure}

In \autoref{fig:timescale_sigma}, we show in each PS1 $griz$ band the distribution of flux change of any pair of observations with time intervals of 3$\pm$0.5d,  9$\pm$0.5d, 30$\pm$0.5d, and 365$\pm$0.5d. The standard deviation of the distribution for each time interval, which measures the amplitude of variability for the given timescale, is indicated by $\sigma_{3}$, $\sigma_{9}$, $\sigma_{30}$, and $\sigma_{365}$ in \autoref{fig:timescale_sigma}. From the distribution, we find the variability in $g$ band being more sensitive to timescale than the other 3 filters.

The measurement uncertainty given by our pipeline is comparable to $\sigma_{3}$ in each PS1 band. Although 3 objects in our sample have interband observation gap as large as 7-9 days, we see in \autoref{fig:timescale_sigma} that $\sigma_{9}$ is only $\sim$36\% greater than $\sigma_{3}$. We find the difference insignificant compared to the uncertainties introduced by subtracting against a bright host galaxy, which can increase the measurement error by a factor of 2.

\subsection{Galactic extinction correction}

We correct for Galactic extinction before performing model fitting to our
difference spectra. For \textsl{GALEX} $NUV$ data, we use the band extinction
given in \cite{2007ApJS..173..293W} that is computed by assuming the
\cite{1989ApJ...345..245C} extinction law and R$_V$ = 3.1.

\begin{equation}\label{eq:extinctuv}
A(NUV)/E(B - V ) = 8.2
\end{equation}

On the other hand, the Galactic extinction of each PS1 band are
computed as a function of PS1 stellar color with
\cite{1999PASP..111...63F} extinction law using R$_V$ = 3.1 by
\cite{2012ApJ...750...99T}. The Galactic extinction of PS1 bands are
given as follows:

\begin{eqnarray} \label{eq:extinctopt}
\begin{aligned}
& A(g_{\rm P1})/E(B - V ) = 3.613 - 0.0972(g_{\rm P1} - i_{\rm P1})+0.0100(g_{\rm P1} - i_{\rm P1})^2 \\
& A(r_{\rm P1})/E(B - V ) = 2.585 - 0.0315(g_{\rm P1} - i_{\rm P1})\\
& A(i_{\rm P1})/E(B - V ) = 1.908 - 0.0152(g_{\rm P1} - i_{\rm P1})\\
& A(z_{\rm P1})/E(B - V ) = 1.499 - 0.0023(g_{\rm P1} - i_{\rm P1})
\end{aligned}
\end{eqnarray}

We use the E(B-V) value from \cite{1998ApJ...500..525S} dust map together with
\autoref{eq:extinctuv} and \autoref{eq:extinctopt} to correct for the Galactic
extinction in our difference spectra.

\subsection{Fitting the difference-flux SEDs}
\label{subsec:application}

We use \autoref{eq:model_tay} to fit the observed $\Delta f$SEDs in the
rest frame. The partial derivative may be further simplified to

\begin{equation} \label{eq:model_sim}
\left. \partial g_\lambda \over \partial T^* \right|_{\bar{T}^*} =  {{\pi \, 2 h^2 c^3} \over {\lambda^6 k \bar{T^*}^2}} \int\displaylimits_{1}^{\infty} { {t(s) } \over \left({\exp \left( h c \over \lambda \, k \bar{T}^* t(s) \right) - 1 } \right)^2} \, 4 \pi s \, ds
\end{equation}

Since the shape of the difference spectrum is only affected by the integral,
the parameters in this model can be further reduced to one free parameter
$\bar{T}^*$ with a normalization factor C = $\sqrt{\sum{y_j
y_{\text{model},j}}/\sum{y_{\text{model},j}^2}}$, where $y_j$ is the observed
flux difference in band $j$ and $y_{\text{model},j}$ is the flux difference
predicted by the model in band $j$ given $\bar{T}^*$. The code integrates 
\autoref{eq:model_sim} numerically from 6$r_{g}$ out to 300 $r_{g}$.

We adopt both the P06 model and a model-independent power-law $f_\lambda
\propto \lambda^{-\alpha_\lambda}$ and perform least squares fitting weighted
by the measurement error on our $\Delta f$SEDs. We calculate the
goodness-of-fit ($\chi^2_\nu$) that is defined as:

\begin{equation}
\chi^2_\nu = \frac{1}{N-2}\sum^{N}_{j}\left(\frac{y_j-C \cdot y_{\text{model},j}}{\sigma_j}\right)^2,
\end{equation}

where $N-2$ is the number of degrees of freedom of the fit. 
The 90\% confidence interval of $\bar{T}^*$ is also calculated by varying $\bar{T}^*$ until $\Delta \chi^2$ = 2.706 if the source has more than 4 data points (two degrees of freedom).

\subsection{Spectral fitting}
\label{subsec:specfit}
We describe in this section the method by which we measure redshift, line widths of H$\beta$ and \MgII, and continuum luminosity from single epoch spectra to derive physical quantities such as black hole mass and accretion rate. All of our spectra are corrected for Galactic extinction using the Milky Way extinction curve from \cite{1989ApJ...345..245C} with $R_v$ = 3.1 and the \cite{1998ApJ...500..525S} dust map.

For H$\beta$ fitting, we first measure the continuum by fitting a linear
function to two $\sim$ 60 \AA\ wide sampling windows on the left of H$\beta$
and on the right of [\OIII] $\lambda \lambda$4959,5007 doublet. We remove the
continuum and fit three Gaussians to the [\OIII] $\lambda \lambda$4959,5007
and the narrow component of H$\beta$ simultaneously. The linewidths
and the redshift of the three Gaussian profiles are forced to change together but
the amplitudes of flux density are allowed to be free. We then use the redshift
determined in this step as the systemic redshift to shift the spectrum back to
its rest frame and measure the narrow linewidth once again. The broad H$\beta$
emission line is also modelled with a single Gaussian that is allowed to have a
velocity offset. We fix the narrow component FWHM and fit both narrow and
broad H$\beta$ and [\OIII] $\lambda \lambda$4959,5007 with four Gaussian
profiles simultaneously.

\begin{figure*}
\centering
\includegraphics[width=7in,angle=0]{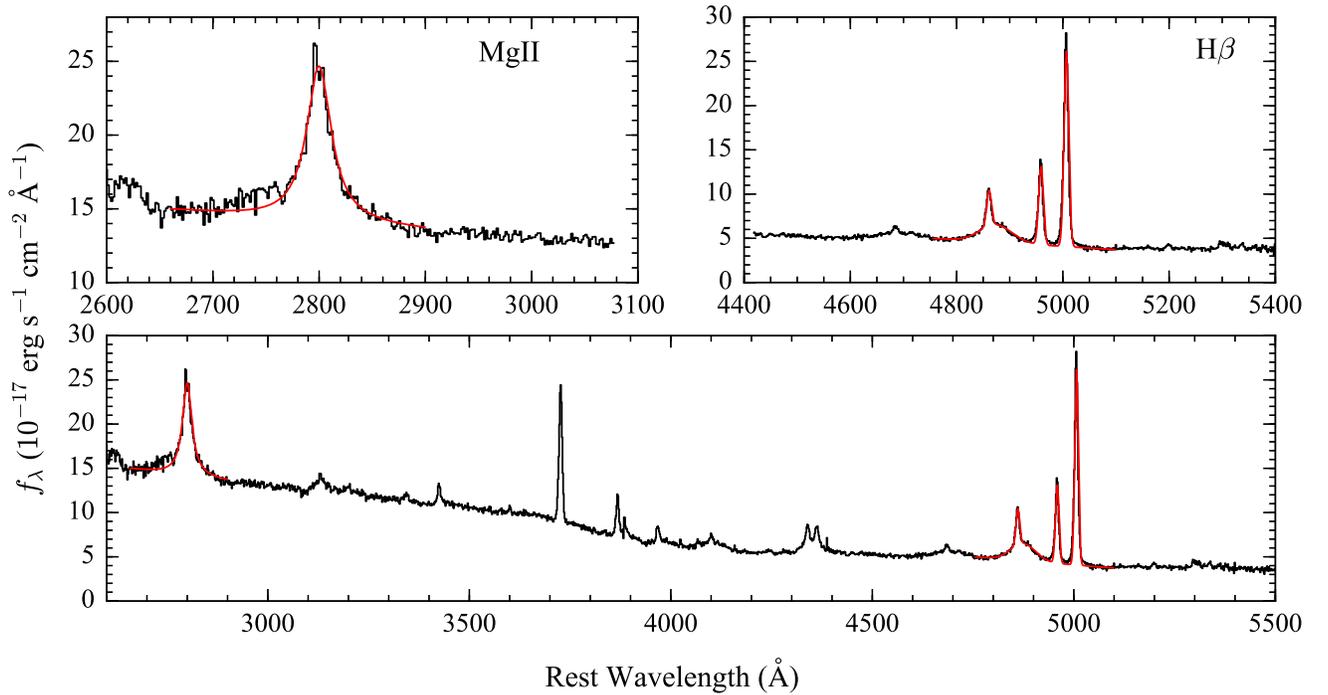}
\caption{The optical spectrum of ELAISN1\_MOS15-17 is shown in black line
while the best-fit model is shown in red line. The upper two panels show the
zoom-in view of \MgII\ and H$\beta$, respectively. The fitting procedure is
described in \autoref{subsec:specfit}.}
\label{fig:spec}
\end{figure*}

\begin{figure}
\includegraphics[width=3.5in, angle=0]{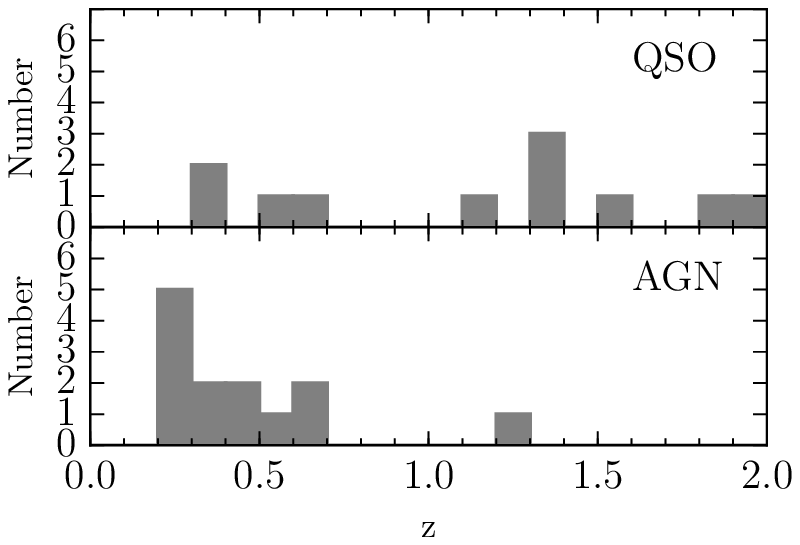}
\caption{The redshift distribution of the AGNs and quasars in our sample. The redshift is
measured as described in \autoref{subsec:specfit}.}
\label{fig:zdis}
\end{figure}

For \MgII, we fit a linear function to the continuum on both sides of \MgII\
and subtract the continuum. We modelled \MgII\ with a single Lorentzian
profile instead of a Gaussian because the \MgII\ lines in our spectra are
better described by the former. We extract the redshift from \MgII\ line
fitting if the object does not have catalogued redshift from SDSS. Note that
we do not treat the \MgII\ line as \MgII\,$\lambda \lambda$ 2796, 2803 doublet
because the double-peaked \MgII\ features in our spectra are unresolved at our
spectral resolution. It is true that by fitting one single line we inevitably account for the narrow \MgII\,$\lambda \lambda$ 2796, 2803 contamination. Nevertheless,
\cite{2011ApJS..194...45S} compared the broad \MgII\ FWHM measurement from
treating \MgII\ as a single line and a doublet and found the result to be
consistent. An example of the best-fit spectrum is shown in 
\autoref{fig:spec}. The redshift distribution of AGN and quasars are shown in
\autoref{fig:zdis}.

% Shen et al. 2011 DSS data release 7 

\subsection{Accretion rate}

%See the definition for converting differential flux to differential luminosity at http://arxiv.org/pdf/astro-ph/9905116.pdf

\label{subsec:accretion}
The bolometric luminosity ($L_{\rm bol}$) of quasars is usually measured by multiplying a bolometric correction to the luminosity density at 3000 \AA~ or 5100 \AA. However, this method is subject to the accuracy of flux calibration at a single wavelength. To take advantage of our broad wavelength coverage, we instead apply bolometric correction over a broader wavelength range by assuming the mean quasar SED in \cite{2006ApJS..166..470R}.

We first calculate integrated UV/optical luminosity ($L_{UV-opt}$) using photometry from \textsl{GALEX} $NUV$ \cite{Gezari13}, CFHT $u$ band, and $griz$ from PS1 \citep{2016ApJ...821...86H}. Note that since AGNs suffer from host light contamination, we do not try to derive their bolometric luminosity and the accretion rate.
Then we scale $L_{UV-opt}$ to $L_{\rm bol}$ using the mean quasar SED in \cite{2006ApJS..166..470R}. \cite{2006ApJS..166..470R} constructed for a sample of 259 quasars detected by optical and MIR selection. Photometry data used in the construction of mean quasar SED are SDSS and the four bands of the Spitzer Infrared Array Camera were supplemented by near-IR, \textsl{GALEX}, VLA, and ROSAT data.

\begin{figure}
\centering
\includegraphics[width=3.5in,angle=0]{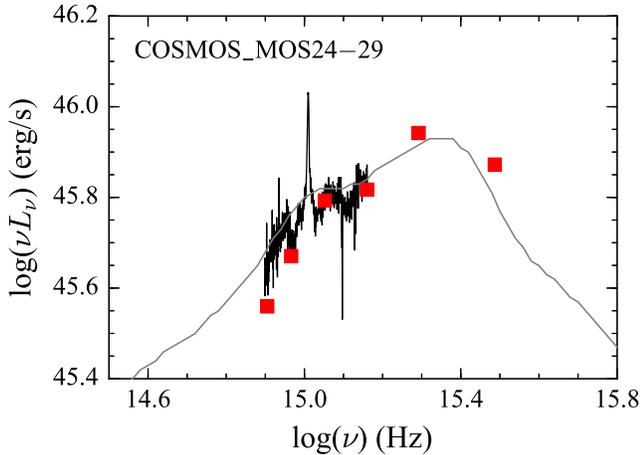}
\caption{SED of a quasar in the sample. The red squares mark the luminosity transformed from \textsl{GALEX} $NUV$ \citep{Gezari13} and the deep-stack $ugriz$ photometry from \cite{2016ApJ...821...86H} in rest frame. The black solid line is the spectrum of the source and the grey solid line is the mean quasar SED. Both grey and black lines are normalized so that the integrated luminosity matches that of the broadband photometry.}
\label{fig:SEDs}
\end{figure}

We convert the flux density $f_\nu$ from broadband photometry to luminosity $L_\nu$ using the following relation
\begin{equation}
\label{eq:fluxtolum}
L_{(1+z)\nu} = \frac{f_\nu}{1+z} \times 4\pi D_L^2,
\end{equation}

\noindent where $\nu$ is the observed frequency and $D_L$ is luminosity distance. An example of one of our quasar UV/optical SEDs is shown in \autoref{fig:SEDs}. The red squares shows the broadband photometry transformed into $L_\nu$. The grey solid line is the mean quasar SED from \cite{2006ApJS..166..470R} normalized to enclose the same $\int L_\nu d\nu$ area as the red squares. 
We then interpolate $\nu L_\nu$ linearly between the effective frequency of $NUV$ and $z$ band with a spacing in $\Delta log\nu$ of 0.05 dex and integrate $\int \nu L_\nu dlog\nu$ in rest frequency to obtain $L_{UV-opt}$. The fractional contribution of $L_{UV-opt}$ to $L_{bol}$ is calculated using the mean quasar SED in \cite{2006ApJS..166..470R}. Finally, we use this fraction to infer $L_{bol}$ from $L_{UV-opt}$. One thing to bear in mind is that the SEDs of quasars show significant scatter \citep{2006ApJS..166..470R}. Therefore, the value derived should be viewed in an average sense. 
The accretion rates
are then calculated by assuming an accretion efficiency of $\eta \sim$0.06 for a
non-spinning black hole, where $L_{\rm bol} = \eta \dot{M}_{\rm accr} c^2$. The result is shown in \autoref{tab:spec}. 

\begin{figure*}
\centering
\includegraphics[width=7in,angle=0]{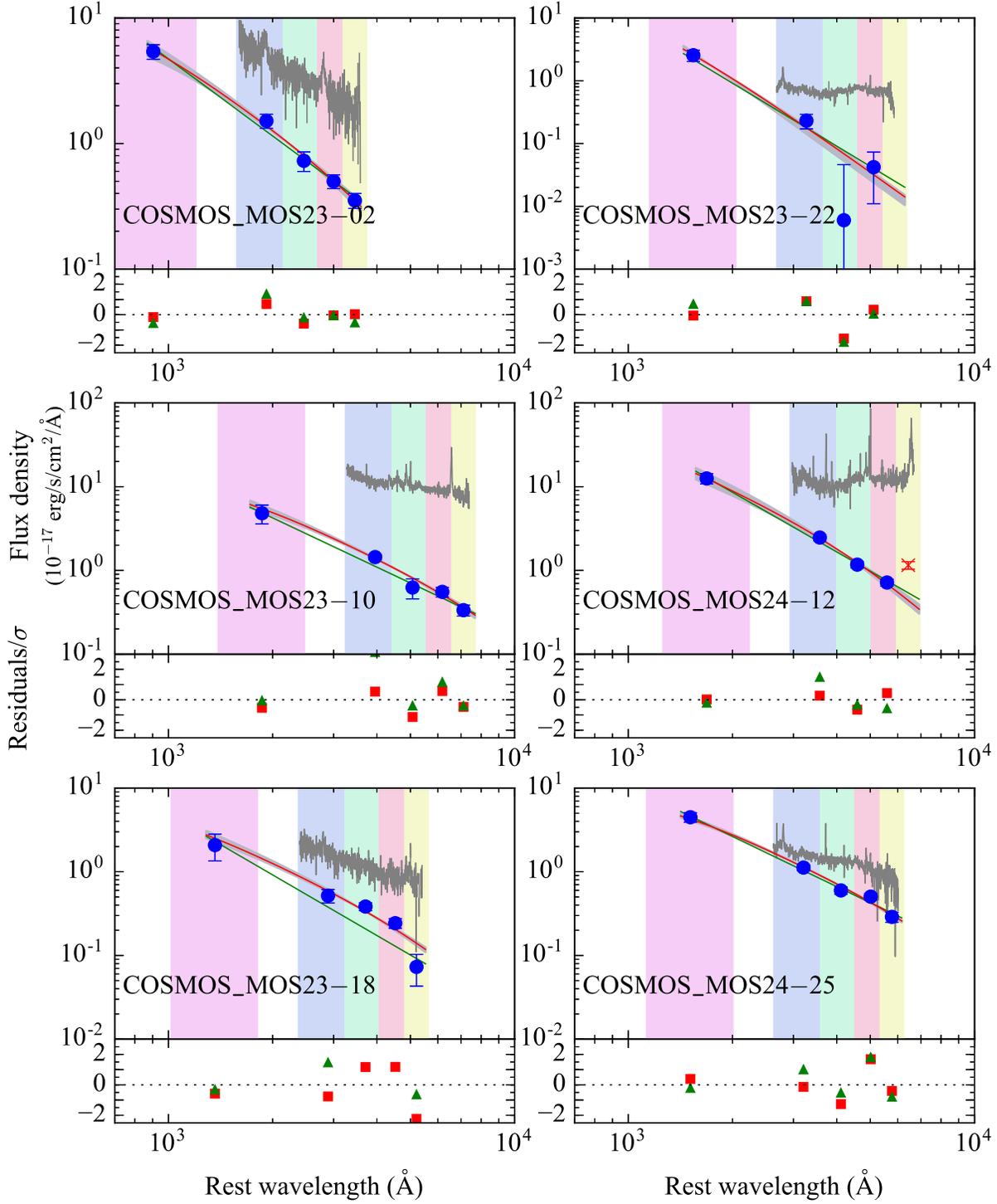}

\caption{Difference-flux SEDs and the best fit results of the 17/23 sources with $\chi_{\nu}^2$ < 3 in the sample. In the upper panel of each subplot, the red line shows the best-fit of P06 model and green line shows the
best-fit of a simple power-law. The horizontal axis is the rest wavelength in \AA~while the vertical axis is the difference-flux on a logarithmic
scale. The shaded area is bounded by two curves corresponding to the 1 $\sigma$ upper
(steep) and lower (flat) limits of $\bar{T}^*$. The filter transmission curves for $NUV,g,r,i,z$ shifted into the rest frame are colored in the background. The optical spectra are also plotted in grey for each source in logarithmic scale. The lower panel shows the residuals of the model fit divided by the uncertainty of each measurement, where the red squares are from the P06 model and the green triangles are from a power-law.}
\label{fig:bestfit_good}
\end{figure*}

\begin{figure*}
\ContinuedFloat
\centering

\includegraphics[width=7in,angle=0]{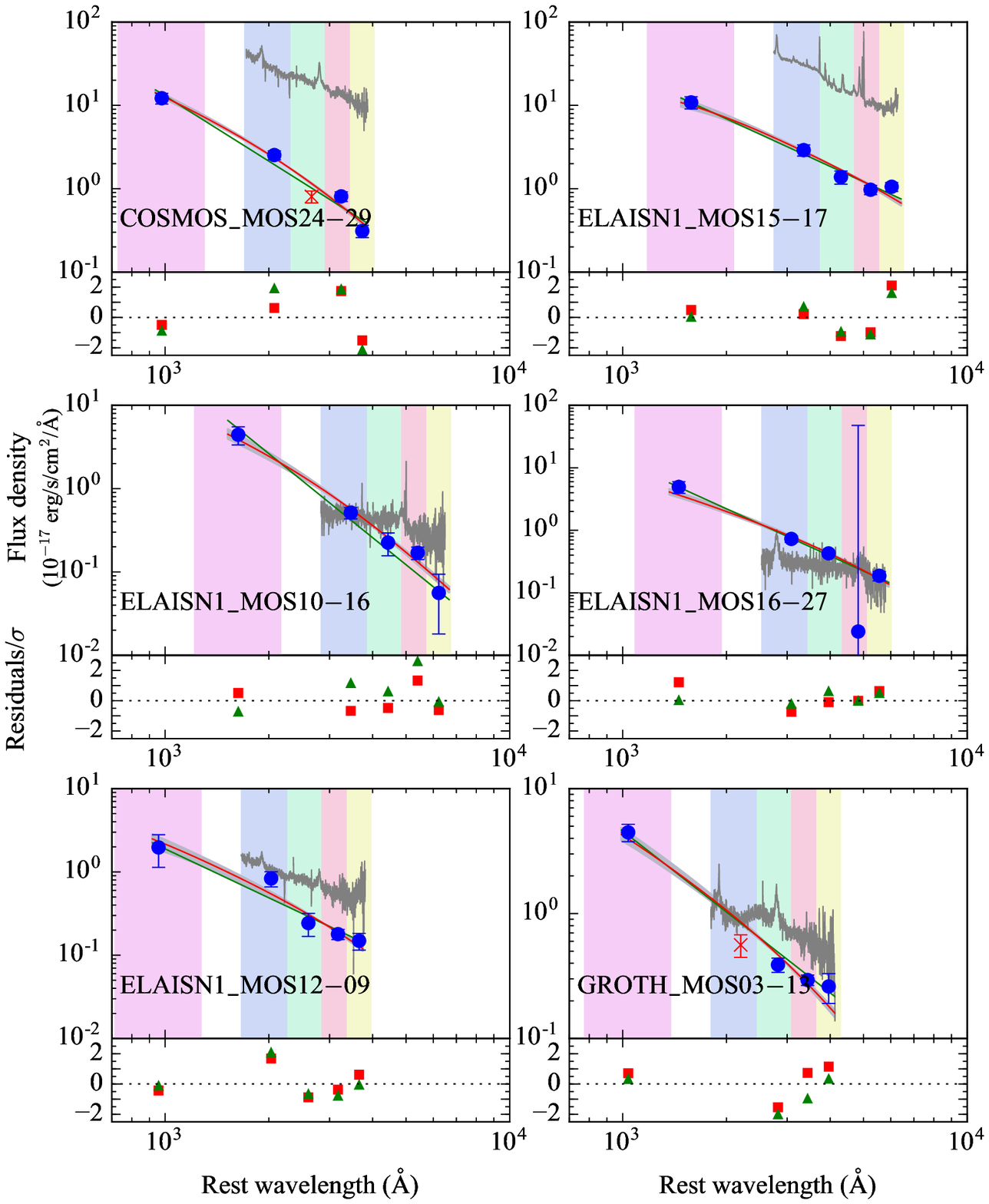}
\caption{Continued.}
\end{figure*}

\begin{figure*}
\ContinuedFloat
\centering

\includegraphics[width=7in,angle=0]{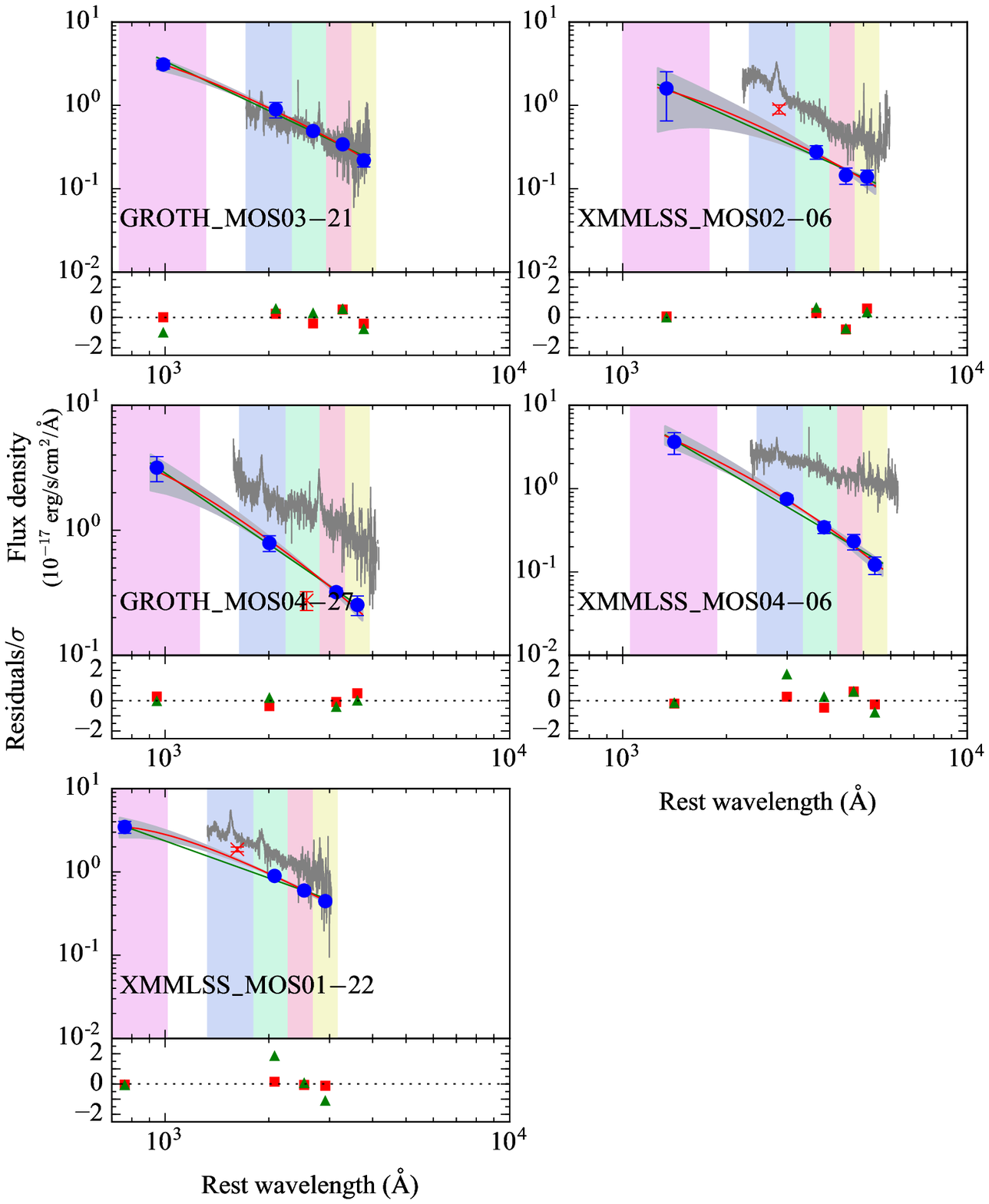}
\caption{Continued.}

\end{figure*}

\subsection{Virial mass}
\label{subsec:bhmass}

The reverberation mapping studies in the optical have yielded direct
measurements of the sizes of the broad-line region ($R_{\rm BLR}$) as a
function of continuum luminosity \citep[e.g.]{2006ApJ...651..775B}. With one
single-epoch spectrum, the black hole mass can be estimated with a measurement
of $L_{\rm cont}$ and the broad-line FWHM, assuming that FWHM reflects
the virialized motion of the line-emitting material. In general, the virial
mass has the form:

\begin{equation}
\frac{M_{\text{bh}}}{M_\odot}= \textit{A} \left[\frac{\lambda L_{\lambda}}{10^{46} \text{erg s}^{-1}}\right]^{\textit{B}} \left[\frac{\text{FWHM}}{\text{km~s}^{-1}}\right]^2,
\end{equation}
where $A$ and $B$ are derived from empirical calibration against AGNs and
$\lambda$ is 5100~\AA and 3000~\AA for H$\beta$ and \MgII, respectively.

We adopt the calibration of \cite{Trakhtenbrot12}. For H$\beta$, A=105 and
B=0.61. For \MgII, A=5.6 and B=0.62. During spectral fitting, we measured
redshift, FWHM of H$\beta$ and \MgII, $f_{5100}$, and $f_{3000}$ in the rest
frame. We convert the flux density to the luminosity density using \autoref{eq:fluxtolum}.

The object class, the accretion rate, the black hole mass, the Eddington
ratio, the FWHM of \MgII\ and H$\beta$ of our sample are listed in 
\autoref{tab:spec}.

\section{Results}
\label{sec:results}

We have compiled a sample of 24 sources that have a variability amplitude greater than 3$\sigma$ in $g$ band with well-aligned epochs across $NUV$ and optical bands. One of the 24 sources, COSMOS\_MOS27-07, shows the spectrum of an early-type galaxy but lacks emission lines typical of AGN so we do not include this source in our model fits (see \autoref{subsec:exception} for discussion of the nature of this source). For the 23 active galaxies, we fit the difference
flux SEDs following the method described in \autoref{subsec:application}.

\begin{figure*}
\centering
\includegraphics[width=7in,angle=0]{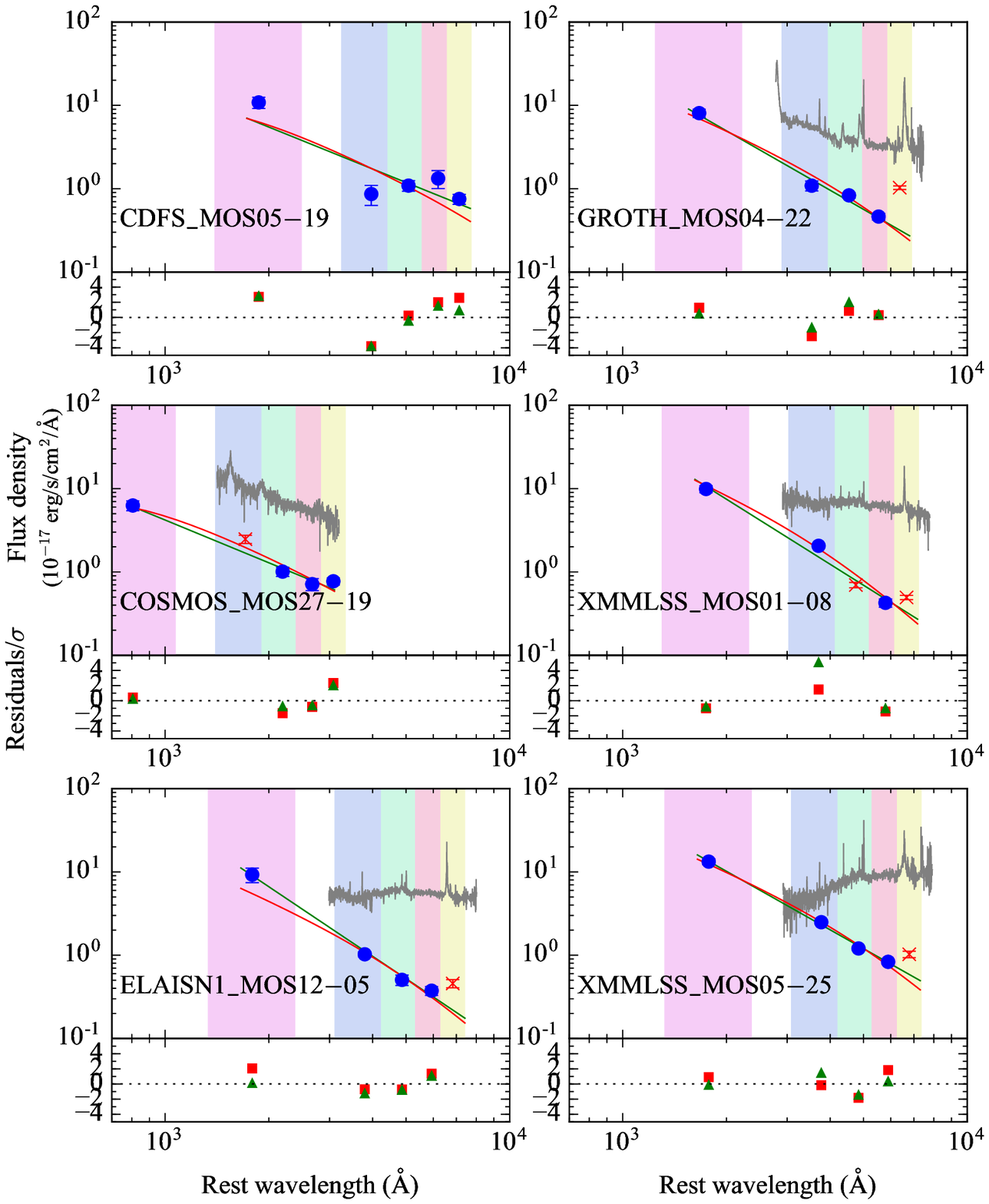}
\caption{Difference-flux SEDs and the best fit results of the 6/23 sources that are poorly-fitted by the model ($\chi_{\nu}^2$ > 3). The notations are the same as in \autoref{fig:bestfit_good}.}
\label{fig:bad_fit}
\end{figure*}

During the model fitting process, we noticed that the bands deviating the most from the P06 model tend to have strong broad emission lines within them. The contribution of broad emission line variability is discussed in \autoref{sec:emissionlines}. We suspect these bands may be contaminated by the emission from the BLR region and could affect our model fitting since the model only accounts for the continuum variability that arises from the emission of the disk. In light of the potential contamination, we repeat model fitting by masking the bands associated with broad emission lines for sources with an original $\chi^2_\nu > 3.0$. Five sources show good agreement with the P06 model ($\chi^2_\nu$ < 3.0) after we mask the broad emission line affected bands while the rest 6 sources still have $\chi^2_\nu > 3.0$. The masked bands are shown as red crosses in \autoref{fig:bestfit_good} and \autoref{fig:bad_fit}. The
summary of the best-fit parameters can be found in \autoref{tab:results}.

We show the $\Delta f$SED and the best-fit models for each well-fitted ($\chi^2_\nu \lesssim 3.0$) source and sources that exhibit deviation from the model ($\chi^2_\nu >3.0$) in \autoref{fig:bestfit_good} and \autoref{fig:bad_fit}. In the upper panel, the vertical axis is the difference-flux in flux density units in logarithmic scale
while the horizontal axis is the rest-frame wavelength in \AA. The red line
shows the best-fit result from the standard thin disk model while the green line shows the best-fit result of a power-law. The shaded area is bounded by two curves corresponding to the upper
(steep) and lower (flat) limit of $\bar{T}^*$ and is only plotted for objects with $\chi^2_\nu \lesssim 3.0$. We also overplot the spectra for each source in grey in \autoref{fig:bestfit_good} and \autoref{fig:bad_fit} in logarithmic scale. The bottom panel shows the residual divided by the uncertainty ($\sigma$) for each band, where red squares are from the P06 model and the green triangles are from the power-law model.

It is worth noting that, for our sample, the P06 model performs significantly better than the power-law in terms of $\chi^2_\nu$ and the P06 model fits qualitatively well to the data points in \autoref{fig:bad_fit} despite their larger $\chi^2_\nu$ values.

\begin{figure}
\centering
\includegraphics[width=3in, angle = 0 ]{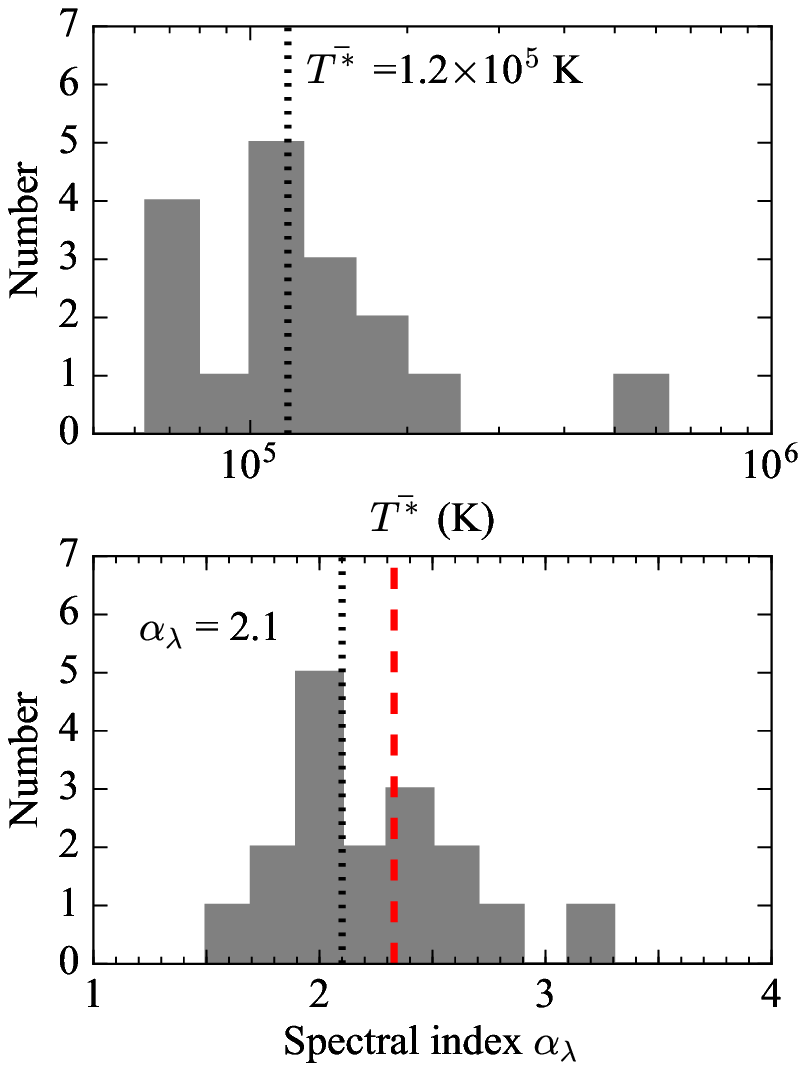}
\caption{Upper panel: The distribution of best-fit mean characteristic
temperature $\bar{T}^*$. Bottom panel: The distribution of spectral index
$\alpha_{\lambda}$. Red dashed line shows the theoretical value of 2.33.}
\label{fig:temppl}
\end{figure}

For the 17 out of 23 objects that are well described by the model, the distribution of the
best-fit mean characteristic temperatures $\bar{T}^*$ is shown in the upper
panel of \autoref{fig:temppl}. The median $\bar{T}^*$ of our sample is
1.2$\times$10$^5$~K, which is higher than the mean characteristic temperature
of 92000~K found for a composite difference spectrum constructed from SDSS
quasars \citep{Pereyra06}. However, the difference composite spectrum in \cite{Pereyra06} is not corrected for Galactic extinction, meaning the underlying difference spectrum may be bluer and therefore could imply a hotter disk temperature than the value derived in \cite{Pereyra06}.

The bottom panel of \autoref{fig:temppl} shows the distribution of best-fit
power law indices. A standard accretion disk model predicts a $f_\nu \propto
\nu^{1/3}$ spectrum for intermediate $\nu$ between the frequencies corresponding to the
innermost disk and the Rayleigh-Jeans limit. This value translates to a
spectral index $\alpha_\lambda$ of $\approx$ 2.33. Although the
bluer-when-brighter trend can be explained as an intrinsic effect of AGN under
the assumption of an accretion disk model, real AGN spectra rarely match this
value. Instead, this spectral index is more often seen toward difference
spectra of AGN. For instance, \cite{Wilhite05} found $\alpha_\lambda$ of 1.35 and
2.00 for the average quasar spectrum and the average quasar difference
spectrum, respectively, using SDSS spectra without correcting for reddening due to Galactic extinction. \cite{Vandenberk01} found a slightly bluer $\alpha_\lambda$ of 1.56 for composite quasar spectrum after correcting for Galactic extinction, but is still not as blue as the theoretical prediction of a thin accretion disk. The composite quasar difference spectrum having a steeper spectral index than the composite quasar spectrum implies that the variable component may come directly from the disk, and that taking the difference flux isolates the radiation from the disk component more cleanly. Our finding of a median of the best-fit power
law index $\alpha_\lambda \approx$ 2.1 is consistent with the spectral index of the SDSS geometric mean composite difference spectrum of quasars in \cite{Ruan14} and is close to the standard thin disk value of 2.33.%see high energy textbook p.460 for the plot

\begin{figure}
\centering
\includegraphics[width=3in, angle = 0]{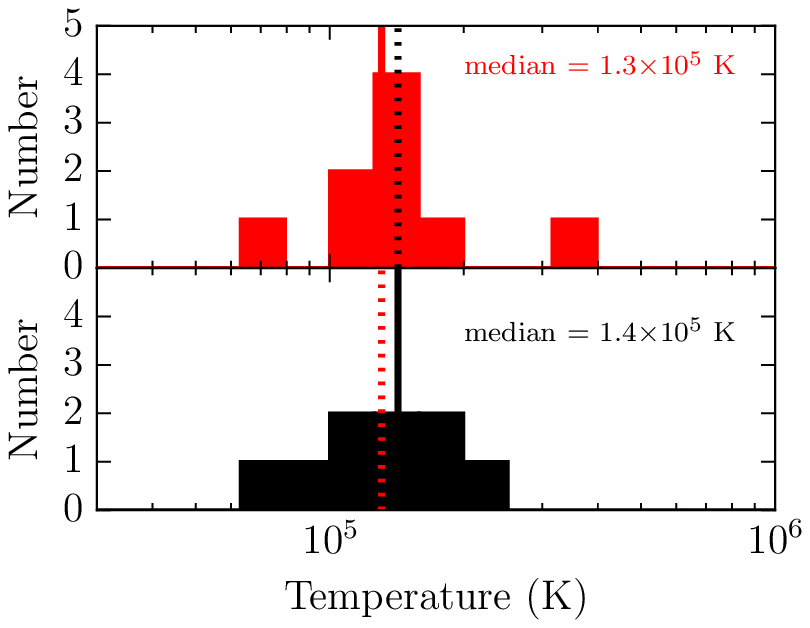} %updated
\caption{The distribution of $\bar{T}^*$ (black) and $T^*$ (red), where
$\bar{T}^*$ is calculated from fitting the $\Delta f$SEDs to the P06
model and $T^*$ is calculated from substituting accretion rate and black hole
mass into \autoref{eq:Tch}. Note that we only include the 12 sources that are
classified as quasars to avoid accounting for galaxy light contamination. Although the two histograms seem to have similar temperature distribution, there is no clear trend in the scatter plot of $\bar{T}^*$ vs $T^*$.}
\label{fig:tcomp}
\end{figure}

The classic thin-disk model predicts the characteristic temperature $T^*$ as a
function of accretion rate and black hole mass (see \autoref{eq:Tch}). For
example, for a quasar accreting at the rate of 1 M$_\odot$/yr with a mass of
10$^9$ M$_\odot$, a standard accretion disk predicts a characteristic
temperature of 70,000~K. As a sanity check, we also calculate the
characteristic temperature $T^*$ using the accretion rates and the black hole
masses derived in \autoref{subsec:accretion} and
\autoref{subsec:bhmass} using photometry and single-epoch spectroscopy. The distribution of $T^*$ and $\bar{T}^*$ are plotted in
\autoref{fig:tcomp}. The data are binned into 0.1 dex intervals. Note that
we only plotted $\bar{T}^*$ for point-like quasars, to avoid estimating the
accretion rates for extended sources, which will have systematic uncertainties in $L_{\rm bol}$ due to host light contamination.
Although we do not see a one to one correlation in the characteristic
temperature $T^*$ and $\bar{T}^*$, the distribution of $T^*$ and $\bar{T}^*$
are consistent with each other, and therefore follow the expectations for an accretion
disk with the ranges of $M_{\rm bh}$ and $\dot{M}_{\rm accr}$ inferred for the quasars in our sample. 

\section{Discussion}
\label{sec:discuss}
\subsection{Emission-line variability}
\label{sec:emissionlines}
Before masking the bands containing broad emission lines, 11/23 sources in \autoref{tab:results} have large $\chi^2_\nu$ values and are clearly ill-fitted by the change in accretion rate model. One likely explanation is that the the $\Delta f$SEDs of these objects are contaminated by emission line variability. Broad emission lines are known to both vary in flux and in profile \citep[e.g.]{1990ApJ...354..446W,0004-637X-475-1-106}. In extreme cases such as a 'changing-look' AGN that involves the sudden appearance or disappearance of the broad Balmer lines on the timescale of years \citep[e.g.][]{2014ApJ...788...48S}, the broad band flux may experience a dramatic change due to the change in emission line flux. Using XMMLSS\_MOS01-22 as an example, the difference-flux in $g$ band deviates from the model prediction by $\sim$30\% (\autoref{fig:bestfit_good}), which would inevitably result in a poor model fit if we force to fit all bands. In fact, \cite{Wilhite05} also reported that broad emission lines could vary by 30\% as much as the continuum in the spectroscopic variability study of SDSS quasars. It is important to recognize that the amount of flux change contributed by a broad AGN emission line can be non-negligible compared to the change in continuum in broadband.

We find that masking the broad-emission-line contaminated bands can improve the poor model fit. Although only 5/11 originally ill-fitted sources have $\chi^2_\nu < $ 3.0 after applying band masks, most of the sources have a qualitatively good model fit even when having $\chi^2_\nu > $ 3.0 (\autoref{fig:bad_fit}).

\subsection{COSMOS\_MOS27-07}
\label{subsec:exception}

\begin{figure*}
\centering
\includegraphics[width=5in, angle=0]{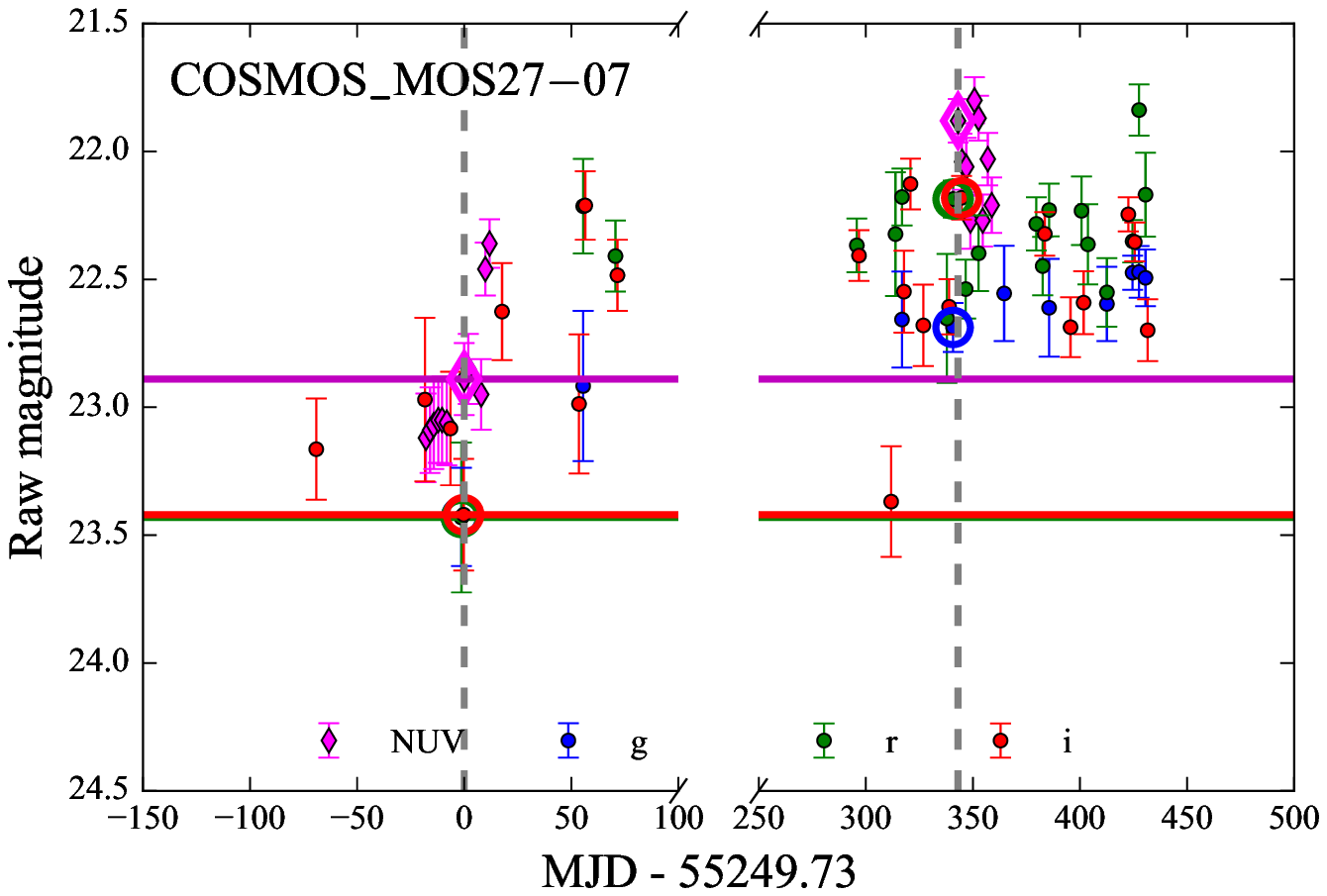}
\caption{The light curve of COSMOS\_MOS27-07 with epochs shown in time with respect to the
reference epoch. The notations are the same as defined in \autoref{fig:lc}.}
\label{fig:lc_tde}
\end{figure*}

While examining the spectra, we found one source, COSMOS\_MOS27-07, that lacks common AGN emission lines even though classified as an AGN from its morphology and light curve. The light curve of
COSMOS\_MOS27-07 is shown in \autoref{fig:lc_tde}.

COSMOS\_MOS27-07 resides in an early-type host galaxy with colors $u-g = 1.71$ and
$g-r = 1.64$. The \textsl{GALEX} TDS catalog reports a maximum variability
amplitude ($\Delta m_{max}$) of \textgreater 1.37 mag, with significant
variability on timescales of $1-2$ years \citep{Gezari13}. It is unlikely that
such amplitude of variability in $NUV$ is caused by core-collapse supernovae
since their $NUV$ light curves are powered by expanding shock-heated ejecta,
which cool quickly with time. 

Such variability can be powered by a
tidal disruption event (TDE), which features persistent emission in the UV ($\gtrsim$ 1 year). However, the long term plateau in the GALEX light curve is unlike the power-law decay observed in TDEs. Observations in X-ray wavelengths may reveal if this object is a X-ray bright, optically normal galaxy \citep[XBONG,][]{2002astro.ph..3019C}.

\subsection{Timescales}
\label{subsec:timescales}
Due to the time sampling of \textsl{GALEX} TDS, the $\Delta f$SEDs in our
active galaxy sample probe a variability timescale of about a year. An
assumption in our model is that enough time has elapsed for the the disk to
adjust to a different mass accretion rate. However, the timescale associated
with mass inflow, the viscous timescale, is on the order of 1000 years
assuming a characteristic disk with thickness $h/r = 0.01$, $M = 10^8
M_{\odot}$ and $\alpha = 0.1$ at $r \sim 100 r_g$, where $r_g \equiv 2GM/c^2$ is the gravitational radius. In general, the viscous timescale is
given by:

\begin{equation}
\tau = \left(\frac{h}{r}\right)^{-2}\frac{1}{\alpha \Omega},
\end{equation}

where $h/r$ is the thickness of the disk in units of the radius, $\alpha$ the
viscosity, and $\Omega$ the angular velocity. To match the timescale we
observed, the mass accretion perturbation must originate from a radius close
to the UV/optical emitting region or have a much thicker disk ($\tau_{visc}
\sim 10$ yr if $h/r = 0.1$).

As pointed out in \cite{Pereyra06}, the observed months-to-years variability timescales are more
consistent with the sound-speed timescale of an accretion disk. Changes in accretion rate might
result in density and pressure perturbations, which can propagate across a
proportion of the disk as sound waves. The DRW model proposed by \cite{2009ApJ...698..895K} to describe the optical variability seen in quasar light curves, also shows fits with a characteristic timescale of $\sim$200 days that is consistent with the thermal timescale.

Another possibility is that the UV/optical variability is driven by
reprocessing of X-ray photons from the inner disk. Under the assumption of a
classic thin disk, the UV and optical spectrum arise predominately from
spatially separated locations. Therefore, one would expect the time-lag
between different bands to be observed as a result of signal propagation. In
fact, inter-band correlations were found in short timescale variability in NGC
2617 and NGC 5548 \citep{2014ApJ...788...48S,2015arXiv151005648F} as an
evidence supporting the X-ray reprocessing scenario. Following an outburst in
NGC 2617, the galaxy has shown disk emission in the UV/optical lagging the
X-ray by 2 to 3 days \citep{2014ApJ...788...48S}. This time lag is consistent
with the light crossing time from the inner X-ray emitting region to the outer
UV/optical emitting region. Although the UV and optical variability seem to
correlate well with the X-rays on short ($\sim$ day) timescales, previous
studies also reported the presence of pronounced long timescale ($\sim$ year)
variability in the optical but not in the X-rays
\citep{2009MNRAS.394..427B,2014MNRAS.444.1469M}, suggesting an independent
mechanism contributing to the optical variability. 

Unfortunately, with this study, we can not differentiate between the steady-state spectrum of
an X-ray-illuminated source from that of a steady-state accretion disk, since
they have the same radial dependence ($r^{-3}$). Simultaneous X-ray and
UV/optical observations on the light-crossing timescale ($\sim$ days), not available in this
study, would be required to distinguish between these two models.

\subsection{Inhomogeneous disk model}

The main weakness of the P06 model lies in the discrepancy between the observed continuum variability timescales, and the much longer timescales over which an accretion rate can change globally in the disk (see \autoref{subsec:timescales} for discussion).  Inspired by the better agreement with thermal timescales \citep{2009ApJ...698..895K}, \cite{Dexter11} proposed an inhomogeneous disk model in which the disk is separated into multiple zones that each vary on the thermal timescale, with an independent temperature fluctuation conforming to the DRW process. Furthermore, \cite{Dexter11} find that the localized temperature fluctuation model predicts a higher flux at shorter wavelength ($\lambda < 1000$ \AA) than the standard thin disk model, and can better describe the composite HST quasar spectrum from Zhang et al. (1997). 

\cite{Ruan14} fitted a composite {\it relative} variability spectrum, created by dividing the composite difference spectrum by the composite quasar spectrum, in the wavelength range of $\lambda = 1500 - 6000$ \AA. They found that the composite relative variability spectrum is better described by the thermal fluctuation model than the standard thin disk model, assuming a fixed 5\% increase in the mass accretion rate for the standard thin disk model. Unfortunately, we cannot directly compare our results by fitting a "relative variability'' SED, because of emission line contamination in our broadband photometry. However, the motivation for constructing the relative variability spectrum in \cite{Ruan14} was to get rid of the effect of internal AGN extinction.  However, when we measure the spectral indices of the single-epoch optical spectra of our active galaxy sample, we find only 5 out of 23 targets with a spectral index redder than the spectral index of the SDSS quasar composite $\alpha_\lambda \approx$ 1.56 \citep{Vandenberk01} that might indicate the presence of internal extinction. In addition, we suspect that comparing the composite relative variability spectrum to the model difference spectra divided by a low-state disk model, is not a fair comparison, since the low-state disk model, as we discussed in \autoref{sec:results}, is much bluer than the observed composite quasar spectrum.  In sum, we don't think that internal extinction is a significant issue for our sample, and we find a good agreement between the P06 model and our variability SEDs.

Recent work of \cite{2016arXiv160503185C} shows the revised thermal fluctuation model with radius-dependent characteristic timescales also shows noticeable departures from the prediction of a standard thin disk model at $\lambda \aplt 1000$ \AA. However, since the shortest rest wavelength probed in our sample is $\approx$ 800 \AA, we cannot look for this characteristic signature. Our analysis shows that, on an object by object basis, the P06 model is an appropriate description of the individual $\Delta f$SEDs of the AGNs and quasars. We do not find the need to invoke the thermal fluctuation model in most of the cases. However, the NUV flux in ELASN1\_MOS12-05 and GROTH\_MOS03-13 is indeed higher than predicted by the P06 model, introducing localized thermal fluctuations might be able to explain the flux excess in NUV.

\section{Summary}

We analyze the spectral variability of a sample of 24 \textsl{GALEX} TDS $NUV$ variability selected sources with large amplitude optical variability (difference flux S/N$ > 3$ in the $g_{\rm P1}$ band) in the PS1 Medium Deep Survey. We gathered single-epoch spectra for the sample, and found 23 out of 24 sources have emission lines characteristic of an unobscured AGN. The bluer-when-brighter trend is observed across our sample. We have fitted
a standard thin disk model with variable accretion rates as well as a simple
power law to the spectral shape of the observed difference-flux SEDs ($\Delta f$SEDs). We have also
measured accretion rate, black hole mass, and Eddington ratio for our sample using photometric and spectroscopic data.
The mean timescale of UV/optical variability we are probing is one year. We find the change in accretion rate model generally describes the $\Delta f$SEDs better than a simple power law. Our
results can be summarized as follows.

\begin{enumerate}
\item The median of the best-fit mean characteristic temperature $\bar{T}^*$
is 1.2$\times$10$^5$~K. The distribution of $\bar{T}^*$ is consistent with the distribution of the characteristic temperature $T^*$ (median $1.3\times
10^5$ K) obtained from independently measured accretion rates and black hole
masses.
\item The best fit power-law indices $\alpha_{\lambda}$ of the difference
spectra have a median of 2.1, which is consistent with the value
previously found in \cite{Wilhite05} and \cite{Ruan14} and is close to the spectral index
$\alpha_{\lambda}$ $\approx$ 2.33 of a classical thin disk.

\end{enumerate}

Our results suggest that the spectral shape of large-amplitude AGN variability on long timescales ($\sim$ years) is well described by a thin steady-state disk with a variable accretion rate.  However, at the wavelength range and timescales we are probing, we cannot distinguish between this simple variable accretion disk model, and a more complicated disk model with localized thermal fluctuation zones, or X-ray reprocessing.  Broadband monitoring of AGN on the timescale of days from the soft X-rays to optical would help distinguish between these scenarios.

\acknowledgements

We thank the anonymous referee for the valuable comments and suggestions that helped to improve this paper.
S.G. was supported in part by NSF CAREER grant 1454816. Some of the observations reported here were obtained at the MMT Observatory, a joint facility of the Smithsonian Institution and the University of Arizona.  We thank R. Foley for his contribution to the PS1 transients program.  The Pan-STARRS1 Surveys (PS1) have been made possible through contributions of the Institute for Astronomy, the University of Hawaii, the Pan-STARRS Project Office, the Max-Planck Society and its participating institutes, the Max Planck Institute for Astronomy, Heidelberg and the Max Planck Institute for Extraterrestrial Physics, Garching, The Johns Hopkins University, Durham University, the University of Edinburgh, Queen's University Belfast, the Harvard-Smithsonian Center for Astrophysics, the Las Cumbres Observatory Global Telescope Network Incorporated, the National Central University of Taiwan, the Space Telescope Science Institute, the National Aeronautics and Space Administration under Grant No. NNX08AR22G issued through the Planetary Science Division of the NASA Science Mission Directorate, the National Science Foundation under Grant No. AST-1238877, and the University of Maryland.

\bibliography{agn2}

\begin{deluxetable}{lcccccccccc} 
\tablecolumns{11} 
\tablewidth{0pc} 
\tablecaption{Parameters derived from single-epoch spectroscopy} 
\tablehead{
\colhead{\textsl{GALEX} objID} & \colhead{PS1ID}  &\colhead{Class} &\colhead{z}  &\colhead{log($L_{\rm bol}$)\tablenotemark{d}} & \colhead{$\dot{M_{\rm accr}}$}&  \colhead{log10($M_{\rm bh}$)} &\colhead{L/L$_{\rm Edd}$}   & \colhead{FWHM(H$\beta$) } & \colhead{FWHM(MgII) } & \colhead{Telescope}\\\colhead{} & \colhead{}  &\colhead{} &\colhead{}  &\colhead{erg s$^{-1}$} &\colhead{$M_\odot$ yr$^{-1}$ } &  \colhead{$M_\odot$ } &\colhead{}  & \colhead{km $s^{-1}$} & \colhead{km s$^{-1}$} & \colhead{}} \\ 
%\cline{2-4} \cline{6-8} \\ 
\startdata 
CDFS\_MOS05-19\tablenotemark{a}    & CfA10I080348  & AGN   & 0.213 & -           & -     & -    & -        & -         & -       & 2DF Galaxy Survey \\
COSMOS\_MOS23-02  & CfA10D030238  & QSO   & 1.507 & 45.83       & 1.96  & 9.06 & 0.05     & -         & 5752.41 & SDSS              \\
COSMOS\_MOS23-10  & CfA10B010116  & AGN   & 0.217 & -           & -     & 7.62 & -        & 3274.5    & -       & SDSS              \\
COSMOS\_MOS23-18  & CfA10L110530 & QSO   & 0.664 & 44.84       & 0.20  & 8.12 & 0.04     & -         & 3614.79 & SDSS              \\
COSMOS\_MOS23-22  & CfA11B130112 & AGN   & 0.471 & -           & -     & 7.51 & -        & -         & 2046.52 & MMT               \\
COSMOS\_MOS24-12  & CfA10C020101  & AGN   & 0.346 & -           & -     & 8.54 & -        & 7164.72   & -       & SDSS              \\
COSMOS\_MOS24-25  & CfA10L110532 & QSO   & 0.500  & 44.55       & 0.10  & 7.13 & 0.22     & -         & 1757.76 & MMT               \\
COSMOS\_MOS24-29  & CfA10D030011  & QSO   & 1.319 & 46.12       & 3.83  & 8.97 & 0.12     & -         & 3848.65 & SDSS              \\
COSMOS\_MOS27-07  & PS1-10uq  & AGN \tablenotemark{b}  & 0.312 & -           & -     & -    & -        & -         & -       & MMT               \\
COSMOS\_MOS27-19  & CfA10A000244 & QSO   & 1.814 & 46.22       & 4.84 & 8.78 & 0.23     & -         & 3293.83 & SDSS              \\
ELAISN1\_MOS10-16 & PS1-10amm  & QSO   & 0.391 & 44.13       & 0.04  & 7.77 & 0.02     & 5369.35   & -       & MMT               \\
ELAISN1\_MOS12-05 & CfA10H070227  & AGN   & 0.267 & -           & -     & 8.18 & -        & 5860.03   & -       & SDSS              \\
ELAISN1\_MOS12-09 & CfA10I080066  & QSO   & 1.369 & 45.32       & 0.60  & 8.31 & 0.09     & -         & 2994.35 & MMT               \\
ELAISN1\_MOS15-17 & CfA10G060134  & AGN   & 0.435 & -           & -     & 8.08 & -        & -         & 3029.09 & MMT               \\
%*ELAISN1\_MOS16-16 & 40526  & AGN \tablenotemark{b}  & 0.212 & -           & -     & -    & -        & -         & -       & SDSS              \\
ELAISN1\_MOS16-27 & CfA10G060158  & AGN   & 0.558 & -           & -     & 7.96 & -        & -         & 6355.37 & MMT               \\
%*ELAISN1\_MOS16-30 & 61100  & QSO   & 1.208 & 45.30       & 0.58  & 9.06 & 0.23     & -         & 7868.88 & MMT               \\
GROTH\_MOS03-13   & CfA10F050930  & QSO   & 1.186 & 45.34       & 0.65  & 8.56 & 0.05     & -         & 5035.06 & MMT               \\
GROTH\_MOS03-21   & CfA10D030207  & AGN   & 1.296 & -           & -     & 7.68 & -        & -         & 2069.92 & MMT               \\
GROTH\_MOS04-22   & CfA10E040161  & QSO   & 0.360 & 44.36       & 0.07  & 7.84 & 0.03     & -         & 3693.61 & SDSS              \\
GROTH\_MOS04-27   & CfA10F050068  & QSO   & 1.397  & 45.43       & 0.79  & 8.37 & 0.10     & -         & 2971.59 & SDSS              \\
XMMLSS\_MOS01-08  & CfA10H071147  & AGN   & 0.299  & -           & -     & 7.89 & -        & 4355.55   & -       & SDSS              \\
XMMLSS\_MOS01-22  & CfA10G061209  & QSO   & 1.970 & 45.73       & 1.55  & 8.23 & 0.26     & -         & 2980.13 & MMT               \\
XMMLSS\_MOS02-06  & PS1-10bet  & AGN   & 0.692 & -           & -     & 8.72 & -        & -         & 6729.49 & SDSS              \\
XMMLSS\_MOS04-06  & CfA10H071018  & AGN   & 0.606 & -           & -     & 7.69 & -        & -         & 2226.38 & SDSS              \\
XMMLSS\_MOS05-25  & CfA10H070606  & AGN   & 0.276 & -           & -     & 0.0 \tablenotemark{c}  & -        & 0.0       & -       & SDSS        
\enddata
\label{tab:spec}
\tablenotetext{a}{2DFGS spectrum is not flux-calibrated.}
\tablenotetext{b}{This source is classified as an AGN because of its extended morphology and stochastically varying light curve. However, this source does not have AGN emission lines in the spectrum. See discussion in \autoref{subsec:exception}}
\tablenotetext{c}{No measurement of black hole mass because this source does not have broad emission lines.}
\tablenotetext{d}{We only calculate bolometric luminosity for point-like quasars since we cannot seperate the host galaxy contribution from the AGN light in photometry.}
\end{deluxetable}

\begin{deluxetable*}{lcccccc}[h]
\tablecolumns{6} 
\tablewidth{0pc} 
\tablecaption{Best-fit parameters of the sample} 
\tablehead{
\colhead{\textsl{GALEX} ID}  & \colhead{$\bar{T}^*$ (10$^5$ K)} &\colhead{$\chi_\nu^2$ \tablenotemark{a}}  &
\colhead{$\alpha_\lambda$} & \colhead{$\chi_\nu^2$\tablenotemark{b}} & \colhead{$\Delta t$ (days) \tablenotemark{d}} }%\cline{2-4} \cline{6-8} \\ 
\startdata 
CDFS\_MOS05-19		& 0.58                   & 10.87 & 1.97$^{+0.18}_{-0.44}$ & 33.6 & 330\\
COSMOS\_MOS23-02		& 1.42$^{+0.28}_{-0.27}$ & 0.28 & 2.04$^{+0.19}_{-0.20}$ & 2.48 & 355\\
COSMOS\_MOS23-10		& 0.67$^{+0.10}_{-0.07}$  & 0.79 & 2.11$^{+0.18}_{-0.28}$ & 5.42 & 347\\
COSMOS\_MOS23-18		& 1.06$^{+0.17}_{-0.12}$ & 2.82 & 1.91$^{+0.48}_{-0.12}$ & 14.11 & 347\\
COSMOS\_MOS23-22		& 5.04$^{+12.6}_{-0.62}$ & 1.64 & 3.26$^{+0.02}_{-0.14}$ & 5.29 & 359\\
COSMOS\_MOS24-12		& 1.25$^{+0.26}_{-0.23}$ & 0.35 & 2.43$^{+0.18}_{-0.22}$ & 2.5 & 355\\
COSMOS\_MOS24-25		& 0.76$^{+0.06}_{-0.04}$ & 1.59 & 1.94$^{+0.20}_{-0.16}$ & 4.75 & 347\\
COSMOS\_MOS24-29		& 2.42$^{+0.19}_{-0.15}$ & 3.00 & 2.52$^{+0.26}_{-0.10}$ & 14.96 & 353\\
COSMOS\_MOS27-07\tablenotemark{c}	& 1.18                   & 5.08 & 2.10 & 14.17 & 343\\
COSMOS\_MOS27-19		& 1.08                   & 4.48 & 1.42$^{+0.22}_{-0.20}$ & 4.84 & 355\\
ELAISN1\_MOS10-16	& 1.89$^{+0.30}_{-0.22}$  & 1.03 & 2.78$^{+0.44}_{-0.32}$ & 2.3 & 355\\
ELAISN1\_MOS12-05	& 1.14                    & 3.57 & 2.66$^{+0.34}_{-0.38}$ & 2.45 & 355\\
ELAISN1\_MOS12-09	& 1.52$^{+0.24}_{-0.36}$ & 1.43 & 2.35$^{+0.14}_{-0.48}$ & 6.42 & 48\\
ELAISN1\_MOS15-17	& 0.67$^{+0.11}_{-0.07}$ & 2.43 & 1.85$^{+0.26}_{-0.22}$ & 5.61 & 306\\
%*ELAISN1\_MOS16-16\tablenotemark{d}	& 0.72                   & 3.52 & -0.63 & 13.41\\
ELAISN1\_MOS16-27	& 1.18$^{+0.18}_{-0.14}$ & 0.82 & 2.42$^{+0.30}_{-0.36}$ & 0.27 & 304\\
%*ELAISN1\_MOS16-30	& 2.06$^{+1.45}_{-1.29}$ & 0.1 & -2.84 & 1.86\\
GROTH\_MOS03-13		& 1.64$^{+0.26}_{-0.19}$ & 2.37 & 2.26$^{+0.24}_{-0.28}$ & 2.03 & 362\\
GROTH\_MOS03-21		& 1.12$^{+0.22}_{-0.20}$  & 0.21 & 1.88$^{+0.16}_{-0.18}$ & 1.60 & 356\\
GROTH\_MOS04-22		& 1.09                   & 4.32 & 2.32$^{+0.18}_{-0.20}$ & 5.09 & 356\\
GROTH\_MOS04-27		& 1.14$^{+0.40}_{-0.32}$ & 0.22 & 1.93$^{+0.27}_{-0.32}$ & 0.11 & 316\\
XMMLSS\_MOS01-08		& 1.41                   & 5.22 & 2.46$^{+0.20}_{-0.08}$ & 22.2 & 331\\
XMMLSS\_MOS01-22		& 0.97$^{+0.18}_{-0.15}$  & 0.02 & 1.58$^{+0.16}_{-0.20}$ & 3.96 & 317\\
XMMLSS\_MOS02-06		& 0.78$^{+0.61}_{-0.38}$  & 0.53 & 1.93$^{+0.50}_{-0.48}$ & 1.08 & 329\\
XMMLSS\_MOS04-06		& 1.54$^{+0.54}_{-0.41}$ & 0.25 & 2.56$^{+0.26}_{-0.34}$ & 2.94 & 325\\
XMMLSS\_MOS05-25		& 1.08                   & 3.73 & 2.35$^{+0.12}_{-0.12}$ & 4.25 & 323

\enddata
\label{tab:results}
\tablecomments{Best-fit parameters of the sample after masking bands with strong broad emission lines. The masked bands are shown as red crosses in \autoref{fig:bestfit_good} and \autoref{fig:bad_fit}.}
\tablenotetext{a}{Reduced $\chi^2$ for the change in accretion rate (P06) model.}
\tablenotetext{b}{Reduced $\chi^2$ for the power-law model.}
\tablenotetext{c}{This source has a quiescent host galaxy. See discussion in \autoref{subsec:exception}.}
\tablenotetext{d}{The time interval between the bright and faint epochs for each object.}

\end{deluxetable*} 

\end{document}